\documentclass[%
reprint,
 amsmath,amssymb,
 aps,
]{revtex4-2}
\usepackage[colorlinks=true,citecolor=black,urlcolor=black,linkcolor=black,bookmarks=false,hypertexnames=true]{hyperref}
\usepackage{multirow}
\usepackage{parskip}
\usepackage{float}
\usepackage[utf8]{inputenc}
\usepackage{siunitx}
\usepackage{graphicx}
\usepackage{dcolumn}
\usepackage{bm}

\usepackage{multirow}
\usepackage{listings}
\usepackage{xcolor}

\definecolor{codegreen}{rgb}{0,0.6,0}
\definecolor{codegray}{rgb}{0.5,0.5,0.5}
\definecolor{codepurple}{rgb}{0.58,0,0.82}
\definecolor{backcolour}{rgb}{0.95,0.95,0.92}

\lstdefinestyle{mystyle}{
    backgroundcolor=\color{backcolour},   
    commentstyle=\color{codegreen},
    keywordstyle=\color{magenta},
    numberstyle=\tiny\color{codegray},
    stringstyle=\color{codepurple},
    basicstyle=\ttfamily\footnotesize,
    breakatwhitespace=false,         
    breaklines=true,                 
    captionpos=b,                    
    keepspaces=true,                 
    numbers=left,                    
    numbersep=10pt,                  
    showspaces=false,                
    showstringspaces=false,
    showtabs=false,                  
    tabsize=2
}

\begin{document}

\title{A simple and low-cost setup for part per billion level frequency stabilization and characterization of red He-Ne laser}

\author{Saurabh Kumar Singh, Avinash Kumar, Pranav R. Shirhatti}
\altaffiliation{
Author to whom correspondence should be addressed.
e-mail: pranavrs@tifrh.res.in}
\affiliation{
Tata Institute of Fundamental Research Hyderabad,
36/P Gopanpally, Hyderabad 500046, Telangana, India}

\begin{abstract}
This work describes the frequency stabilization of a dual longitudinal mode, red (632.8 nm) He-Ne laser, implemented using a low-cost microcontroller and its performance characterization using a simple interferometric method. 
Our studies demonstrate that frequency stability up to 0.42 MHz (3$\sigma$, 17 hours) can be achieved using this set up. This simple and low-cost system can serve as an excellent part per billion level frequency reference for several high resolution spectroscopy based applications.

\end{abstract}
\maketitle

\section{Introduction}

Having a precise and accurate frequency reference is essential for high resolution spectroscopy related applications in order to determine absolute frequency or wavelength. Demands on the level of accuracy and precision, defined as $\Delta f/f$,  can vary over several orders of magnitude depending on the exact nature of application. For example, high resolution atomic spectroscopy measurements involving ultra-cold atoms/ions often require frequency references with accuracy and precision ranging from $10^{-9}$ to $10^{-15}$ \cite{peil_evaluation_2014}. Typically, lasers locked to well known atomic transitions, are used for this purpose.

For several molecular spectroscopy based applications the requirements are relatively less stringent.
As an example, a common scenario arising in spectroscopy experiments carried out in internally cold molecular beams, produced by supersonic jet expansion, is that the observed line width is of the order of a few MHz ($\Delta f/f$ $\sim $ $10^{-9}$) \cite{gough1977infrared}.
This is mainly limited by the residual Doppler broadening, caused by a small transverse velocity distribution of the collimated beam and transit time broadening.
Having a simple, low cost frequency reference with part per billion level stability is of great value in such scenarios.
For example, this can serve as a frequency reference in the so called transfer cavity method \cite{riedle1994stabilization,jackson2018laser}, used for frequency stabilization of lasers operating over a wide range of wavelengths.
Here, a stabilized laser (like He-Ne) acts as a frequency reference for a scanning Fabry Perot cavity, which can then be used to lock another laser operating at a different frequency. 

In this work, we implement the well-known polarization stabilization scheme  \cite{balhorn_frequency_1972,bennett_comments_1973,ciddor_two-mode_1983,eom_frequency_2002} for frequency stabilization of red He-Ne lasers (632.8 nm) using simple and low cost electronic components (see SI-1).
Further, we evaluate its frequency stability by means of an interferometric method, built using readily available off the shelf optical components (see SI-2).
We demonstrate that this setup delivers performance comparable to typical commercially available systems (for example: Excelitas Technologies 32734, Newport N-STP-910, Thorlabs HRS015B), with ppb level frequency stability over several hours timescale.
We have tested this setup over a time span of six months and haven't noticed any degradation in its performance.

In the following sections, the design and implementation of a polarization stabilized He-Ne laser using the interferometer used for its characterization is described.  
Further, we characterize its frequency stability and compare its performance with typical commercially available systems.
Finally, we discuss the possible improvements and potential applications in the context of high resolution spectroscopy experiments. 

\section{Experimental methods}
\subsection{Stabilization scheme for He-Ne laser}

One of the primary consideration for such a setup is selecting an appropriate laser cavity that can be used for frequency stabilization, taking advantage of the orthogonally polarized adjacent longitudinal modes \cite{murakami_frequency_1998,ciddor_two-mode_1983}. 
Doppler broadened gain profile of He-Ne laser spans approximately 1.5 GHz (full width half maximum, FWHM) and using a short cavity with a relatively large free spectral range (FSR) of $\sim$ 1 GHz, will lead to two orthogonally polarized longitudinal modes within the gain profile.
Intensity of these two adjacent modes and the corresponding signals (on a photo detector) $S_1$ and $S_2$, at any given instant of time depend on their position within the gain profile and will change if there is a frequency drift (for example due to thermal expansion/contraction of the cavity).
As a result, the normalized change in intensity can be used to generate an error signal ($\Delta$) for frequency stabilization using the following relation:

\begin{equation} 
    \Delta = \frac{S_1 - S_2}{S_1 + S_2}
    \label{eq: errsig}
\end{equation}

\begin{figure*}[ht!]
\includegraphics[width=1\linewidth]{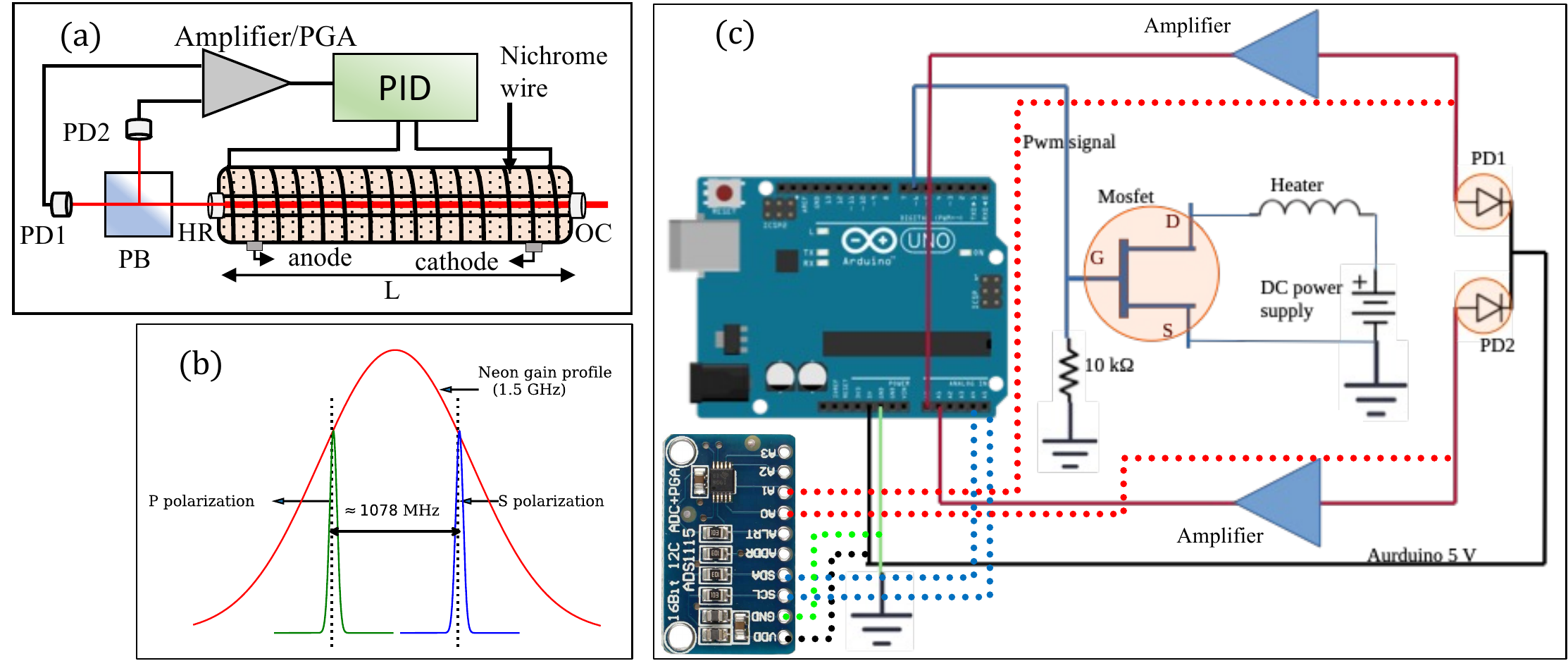}
\caption{(a) Schematic arrangement of all components used in stabilization scheme. PID corresponds to the proportional-integral-derivative feedback system, PD denotes photodiode, PB denotes polarizing beam spliter, PGA denotes programmable gain amplifier, HR and OC denote the high reflectivity end mirror and the output coupler of the He-Ne tube, respectively. (b) Longitudinal mode structure of He-Ne tube. (c) Schematic representation of microcontroller based feedback system used for frequency stabilization. In the first approach (solid lines), an analog amplifier and 10 bit microcontroller are used for frequency stabilization whereas in the second approach (dashed lines), a 16 bit ADC with programmable gain amplifier (PGA) has been used.}
\label{fig:schematic}
\end{figure*}

In this work, a He-Ne laser tube with a cavity length of 139 mm (Melles Griot, 05-LHR-006) was used. 
As per the manufacturer’s data sheet, longitudinal modes in this tube are spaced by 1078 MHz.
Relatively low intensity light beam emerging from the end mirror (opposite to the output coupler) of the He-He tube was used for frequency stabilization. 
This beam was passed through a polarizing beam splitter cube (PBS102 – 10 mm, 620 – 1000 nm, Thorlabs) in order to separate the orthogonally polarized longitudinal modes.
These two separated modes were made incident on two different photodiodes (FDS100 – Si Photodiode, 350 – 1100 nm, Thorlabs). 
Photodiode signals (analog) were amplified (5x gain, non-inverting amplifier using OPAmp uA741C IC)  resulting in a maximum 4 Volts signal in each channel ($S_1$ and $S_2$). These signals were digitized using a low-cost microcontroller (Arduino UNO) with a 10-bit resolution (see figure \ref{fig:schematic}). A second version using a 16 bit ADC which with a programmable gain amplifier (ADS115) was also built and tested.  

Steady state temperature of the He-Ne tube in operation was 60$^\circ$ C (no lock). In order to lock the modes by actively controlling the cavity length, we set the temperature a little higher at 65$^\circ$ C. 
Heating of the tube was achieved by means of a Nichrome wire wrapped around the tube (resistance = 25 ohm) and applying 12 V across it.
Temperature control of the He-Ne tube and hence its cavity length, was achieved using a MOSFET (IRF520N) based switching circuit whose duty cycle was controlled by the same microcontroller.
This was also used to provide the feedback control by means of the well-known proportional-integral-derivative (PID) technique (source code provided in SI-3).

\subsection{\label{sec:level4}Interferometer}

\begin{figure}[H]
  \centering
  \includegraphics[width=8.0 cm]{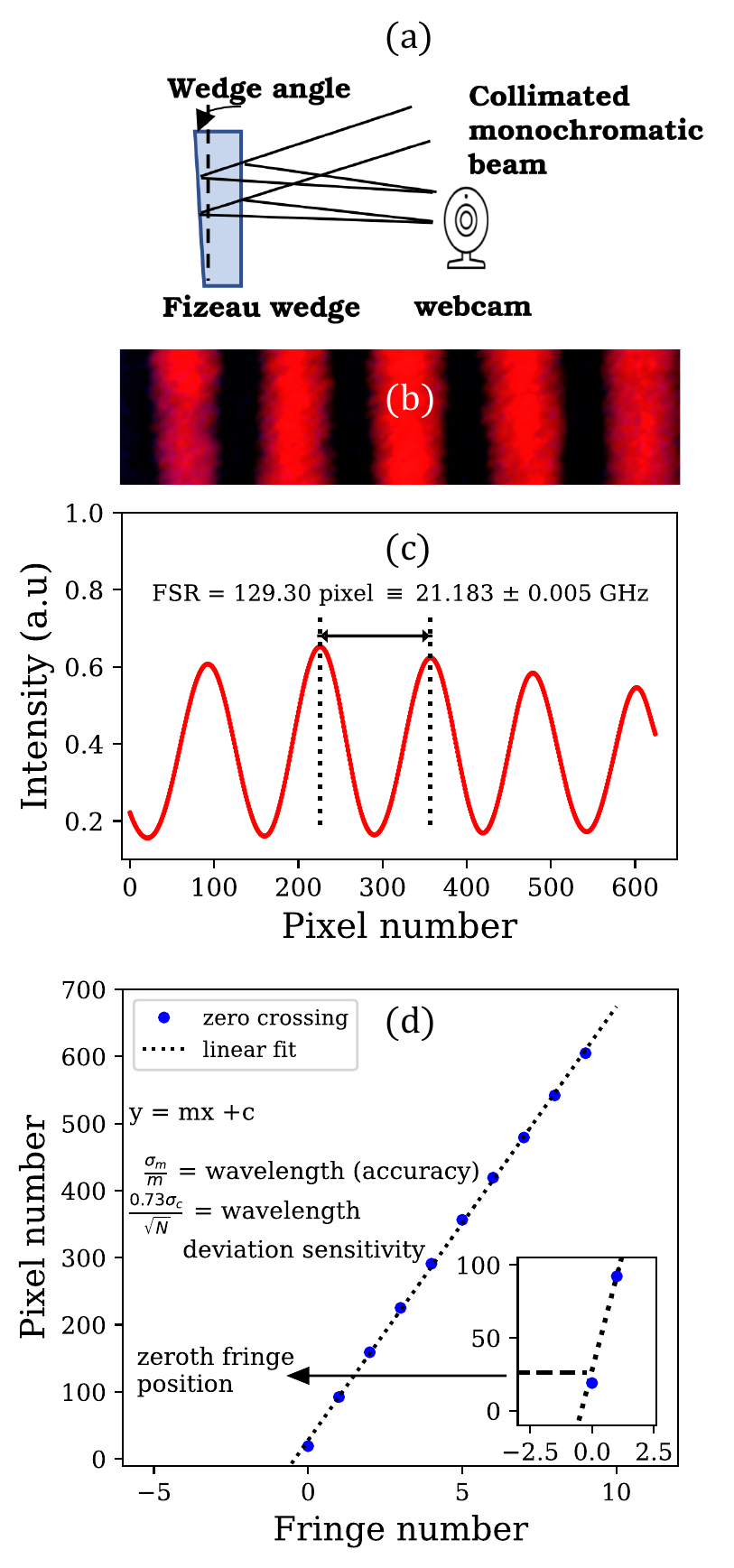}
  \caption{(a) A schematic diagram of our interferometer used in this work for the characterization of the frequency stable He-Ne (b) Interference fringes captured by a webcam (640 $\times$480 pixel), using 632.8 nm input (c) Fringe intensity vs pixel number obtained by summing up the fringe pattern along an axis parallel to the fringe pattern (d) Position of maxima/minima, in terms of pixel number vs the fringe number.}
  \label{fig:fringe_pattern}
\end{figure}

For measuring frequency drifts of the stabilized He-Ne laser we need a reference which has precision better than the He-Ne itself. 
Some options here are: (1) Mixing of the output of our He-Ne laser with an independent stabilized He-Ne laser and monitor its beat spectrum using a spectrum analyzer (2) Using a high precision interferometer such as commercial wavemeters or a stabilized high finesse Fabry Perot cavity to monitor frequency drifts. 
Instead we chose a simpler method which can be readily implemented with off the shelf available components.
In principle, error signal generated by the PID controller itself contains the frequency deviation information. 
However, one needs to first determine the calibration factor using which the change in the error signal can be related to frequency drift. 
In order to do so, we built a simple wavemeter based on a Fizeau interferometer using a 5 mm thick wedge (Thorlabs-WW41050 – Ø1" UVFS Wedged Window, Uncoated, $\leq$ {$\lambda$}/{20} over Central Ø10 mm) and using a low cost webcam sensor (Quantum QHM495LM6 Webcam, 640×480 pixel, lens and IR filter removed). 
A schematic diagram of this set up is shown in figure \ref{fig:fringe_pattern}{(a)}.  
Monochromatic light from the He-Ne laser is coupled through a single mode fiber (Thorlabs - SM, FC/PC, 633-780nm, FT030-Y, 1m) and made incident on the wedge. 
Reflected light from both surfaces forms a characteristic interference pattern which is recorded using a webcam and is shown in figure \ref{fig:fringe_pattern}(b).
A LabView based program was written to analyze this fringe pattern and obtain a plot of intensity vs pixel number as shown in figure \ref{fig:fringe_pattern}(c).
This is analyzed to obtain positions of the maxima/minima of the intensity pattern.
Maxima and minima positions were plotted with fringe numbers and fitted with a straight line to obtain the slope and intercept (see figure \ref{fig:fringe_pattern} (d)).
The intercept corresponds to the zero fringe position (ZFP) and slope corresponds to the distance between two consecutive crests and troughs of the fringe pattern.
In essence, uncertainty in wavelength determination is directly related to the error in determining the slope $\frac{\Delta\lambda}{\lambda}$$=$$\frac{\Delta m}{m}$. Precision (frequency deviation sensitivity) is related to the uncertainty in ZFP which is defined as $\frac{\sigma_{c}}{\sqrt{N}}$ \cite{morris1984fizeau} (${\sigma_{c}}$ is standard deviation in intercept determination from fitting, N is the total number of pixel at horizontal axis). Spacing between two successive crests (or minima), also known as free spectral range (FSR) is defined as c/2nL (n = refractive index, L is the path length of the medium). 
A reasonable estimate of the FSR of the wedge can be simply made by measuring its physical thickness.
Our wedge has length L = 4.8 $\pm$ 0.1 mm, n = 1.45717 for fused silica at 632.8 nm \cite{rodney1954index}, we obtain a value of the FSR to be 21.4 $\pm$ 0.4 GHz. 
A more accurate measurement of FSR was was carried out in different ways listed below (see SI-4 for details). 
In the first method, FSR was obtained using a tunable pulsed dye laser (Lioptec, multimode, linewidth $\sim$ 1.5 GHz, minimum step size $\sim$ 0.1 GHz, Rhodamine B Dye, tuning range 590-605 nm).
Here, the dye laser was scanned and the zeroth fringe position was monitored. 
The frequency change corresponding to change in zeroth fringe position changing by one fringe spacing is equal to the FSR.
Dye laser (DL) was independently calibrated by measuring the well known iodine absorption spectrum \cite{iodine}.
These measurements resulted in a FSR value of 21.1 $\pm$ 0.2 GHz. 
As a second method, we used the fact that the adjacent modes of our He-Ne (orthogonally polarized) are spaced by 1078 MHz (as per the data sheet).
By sending in orthogonally polarized light (using Polarizing beam splitter cube) from the laser one by one to the Fizeau wedge, we observed that the zero fringe position changes by 6.58 pixels. 
Using this relation among pixels and frequency shift we obtain a relation of 1 pixel = 163.83 MHz and the FSR being 21.183 $\pm$ 0.005 GHz. In the first method, the estimated error in the FSR measurement represents an uncertainty in the fitting, whereas in the second method it arose due to the drift of the interferometer during the measurement time interval (see performance of the interferometer section). 
In summary, the FSR was determined to be 21.1 $\pm$ 0.2 GHz, 21.183 $\pm$ 0.005 GHz using two independent methods and these values were found to be consistent with each other.
Using this relation among the ZFP change and FSR [1 pixel = 163.83 MHz, see figure \ref{fig:fringe_pattern}(c)] we are able to quantify the frequency drifts in He-Ne laser. 
These estimations are discussed in detail in SI-4.

\section{Results and discussion}
 \subsection{Locking performance}
\begin{figure}[ht!]
  \centering
  \includegraphics[width=8 cm]{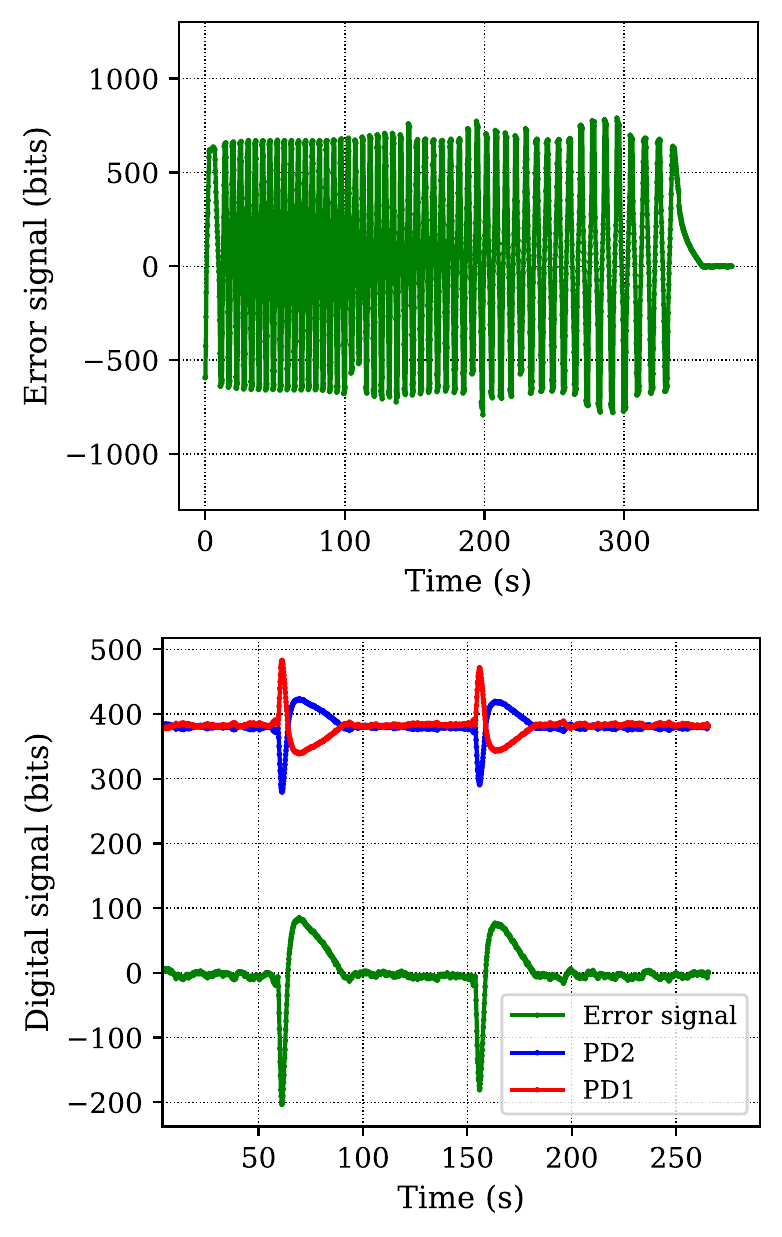}
\caption{(Top) Error signal as a function of time as the laser is started and reaches stable operation, under active feedback control. (Bottom) Response of the stabilized He-Ne to an external disturbance created by blowing cold air for a few seconds (around 60 s and 155 s). Our controller is able to lock the He-Ne within 30-50 s.} 
\label{fig:pid_perf}
\end{figure}

Typically when our He-Ne laser is turned on (initially at room temperature, 298 K), temperature of the tube increases and reaches a steady state in about 15 min, leading to a passively stable operation.
In this duration, the cavity length increases and causes changes in the output frequency, leading to a sweeping behaviour of the longitudinal modes, shown in figure \ref{fig:pid_perf} (top panel).
 
Upon actively controlling the cavity temperature by means of heating using a nichrome wire wrapped around the He-Ne tube, we observed that the time taken to reach steady state is much smaller and our control system can stabilize He-Ne laser within $\sim$ 6 min (see figure \ref{fig:pid_perf}, top panel).
Our feedback control system is reasonably robust to external disturbances, as seen in figure \ref{fig:pid_perf} (bottom panel).
Following a disturbance (jet of cold air blown over the He-Ne tube which causes a  frequency shift of  70-80 MHz approximately), the feedback system locks the laser back again in a matter of 30-50 seconds.  
Having established that our control system is working well and is able to lock the laser, we turn our attention to quantify the frequency fluctuations and drifts observed, in order to evaluate the frequency stability of the He-Ne.

\subsection{Performance of interferometer}
\begin{figure}[ht!]
  \centering
\includegraphics[width=8 cm]{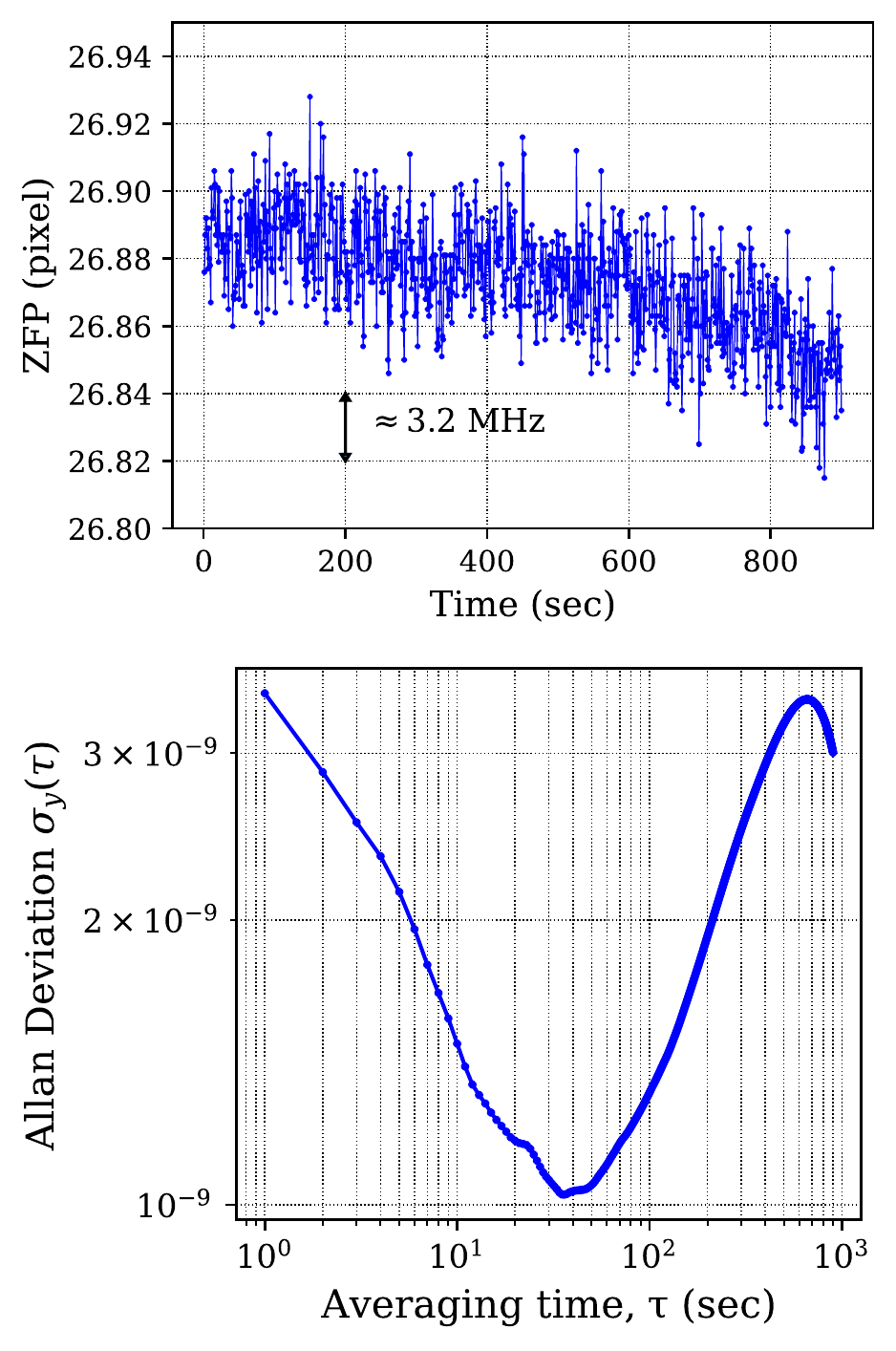}
\caption{(Top) Change in ZPF values (in pixel) measured wrt time for a duration of few hundred seconds (measured at 30 Hz, with averaging of 30 points, resulting in 1 point per sec). This was measured using the stabilized He-Ne as an input. Based on our calibration, 0.02 pixel corresponds to 3.2 MHz. (Bottom) Allan deviation corresponding to the data shown in top panel. In the course of these measurements, the frequency drift in He–Ne was much smaller and could be ruled out based on the monitored error signal over time. Therefore these drifts correspond to the wavemeter itself.} 
\label{fig:wavemeter_stability}
\end{figure}

Our program to analyze the fringes is able to detect changes in fringe position better than $0.02$ pixel, ($\frac{\sigma_{c}}{\sqrt{N}}$) change resulting in ppb level precision.
This can be further increased to sub-ppb level by increasing number of fringes and taking advantage of averaging.
Having said that, care has to be taken to compensate for thermal drifts.
Using low thermal expansion coefficient materials like quartz silica wedge as in our case, one can achieve high precision for a short time ($\sim$15 min, see figure \ref{fig:wavemeter_stability}, top panel).
At longer time scales, effect of thermal drift can be seen. The bottom panel
figure \ref{fig:wavemeter_stability} shows the associated Allan deviation. We conclude that this drift is largely from the wavemeter itself since the He-Ne was locked at a setpoint and the drift in its frequency, determined from changes in error signal, was estimated to be much smaller (relation between error signal and frequency drift is described in the next section).
In principle, this drift can be compensated by measuring the temperature very precisely using a high precision (mK) temperature sensor or by housing the wavemeter in a temperature stabilized housing. 
However, for the present application, this short term stability is good enough to obtain a calibration among the fringe position, corresponding error signal and the frequency drift.
\begin{figure}[ht!]
  \centering
  \includegraphics[width=8cm]{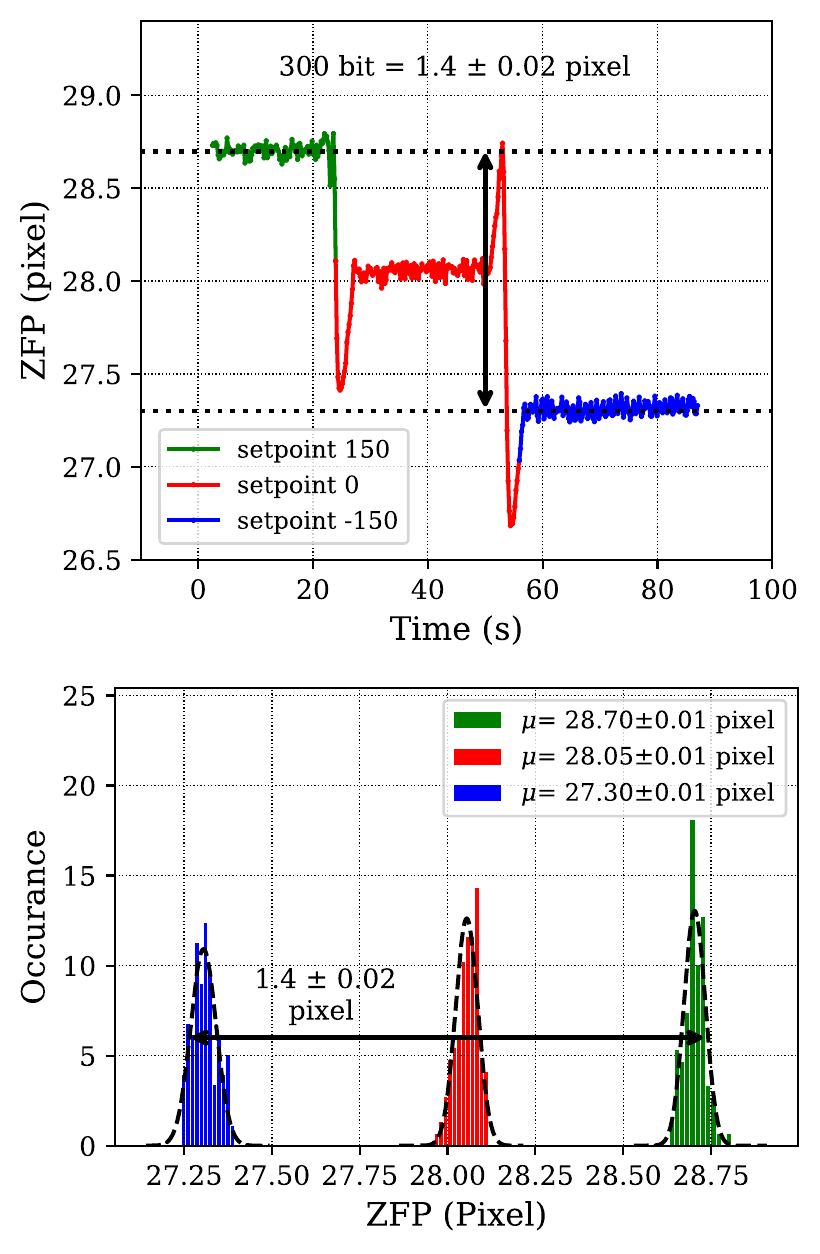}
 \caption{Calibration of error signal (in digital units) to frequency units (in MHz). (Top) Change in ZFP values are plotted with different values of error signal (setpoint). (Bottom) Histogram of the above measurements  showing their distribution and mean values. Based on our calibration (1 pixel = 163.18 MHz), 1 bit corresponds to 0.76 MHz. The uncertainty in mean corresponds to random error along with systematic error (thermal drift in interferometer) as described in SI-4. In the same way, calibration was done for the 16 bit ADC (see SI-6).}
 \label{fig:dig_unit_calibration}
\end{figure}

\subsection{\label{sec:level11} Relation among the frequency change and the error signal}
 
Frequency drift of the He-Ne laser is related to the error signal through the intensity changes in adjacent longitudinal modes, as given by equation \ref{eq: errsig}.
By knowing the precise relationship among error signal (measured in digital units by our detection system) and frequency drift (obtained using the fringe position shifts), we estimate  frequency stability of the locked He-Ne laser.
It is worth pointing out that this method is independent of the long term stability of the wavemeter itself. 
Once the calibration among frequency drift and error signal is established, frequency drift can be estimated from the error signal alone.
\begin{figure}[ht!]
  \centering
  \includegraphics[width=8cm]{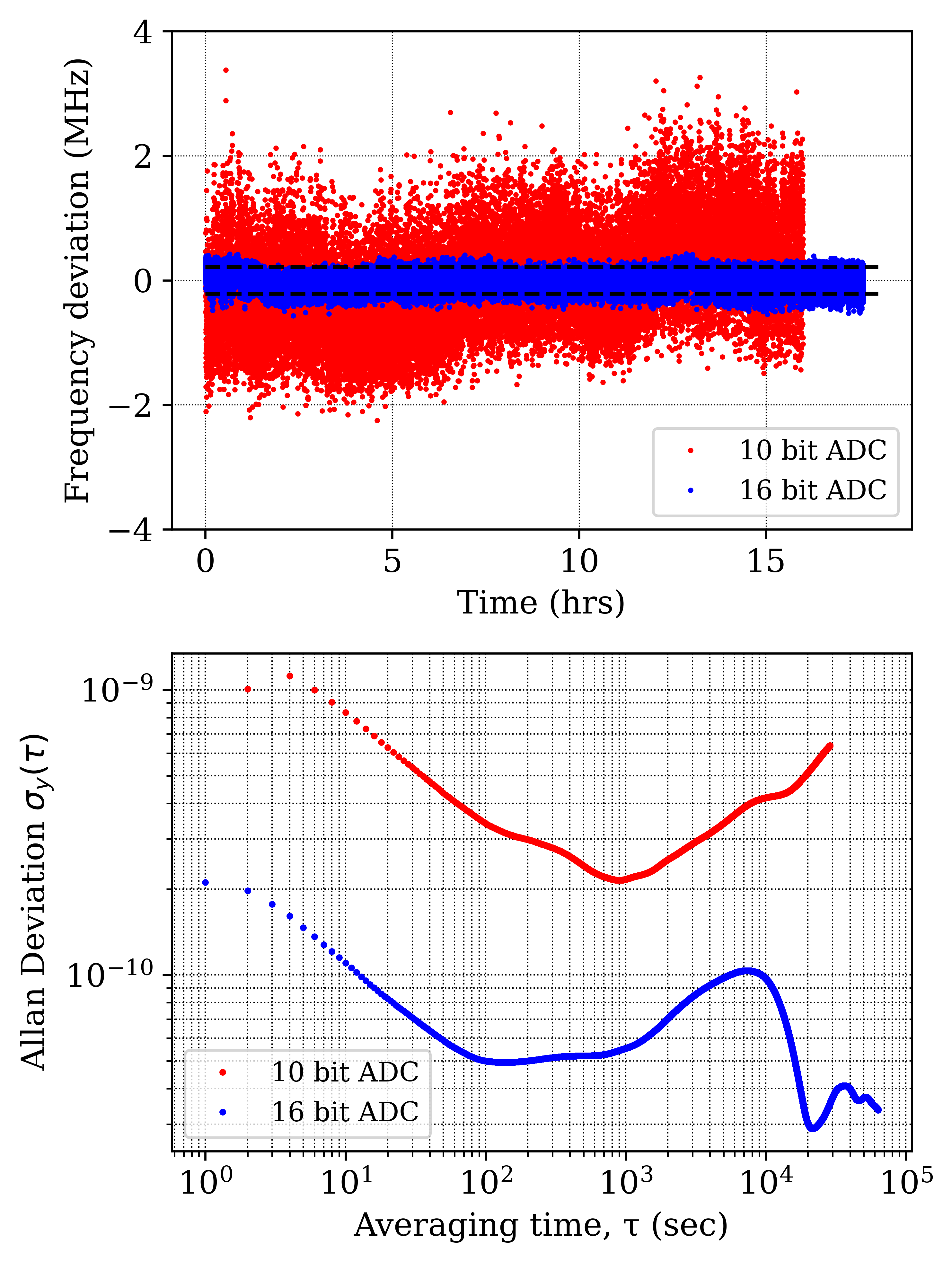}
 \caption{(Top) Frequency deviation in MHz (obtained from change in error signal) measured as a function of time for a duration of 17 hrs. The frequency stability was found to be 0.42 MHz (3$\sigma$) as shown by the two horizontal dashed lines. For the version with 10 bit ADC it was 3 MHz. (Bottom) Allan deviation plot for the above measurements.}
 \label{fig:long_term}
\end{figure}
In order to do this, ZFP (in pixels) was monitored by changing the set-point of PID controller. 
Basically, as we change the set-point (in digital units), frequency of the modes under the gain curve shift and this change was measured using our interferometer.
By knowing the previously obtained relation among fringe position shift and the corresponding frequency change (1 pixel = 163.83 MHz), 
we conclude that one digital unit of error signal corresponds to a frequency change of 0.76 $\pm$ 0.01 MHz (see figure \ref{fig:dig_unit_calibration}). 
To increase the readout resolution we have also used a 16 bit ADC which has a built-in programmable gain amplifier (see figure 1). Using this we found that change in error signal by 1 bit corresponds to 0.008 MHz (see SI-5).

\subsection{\label{sec:level12}  Long term locking stability}
 
We measured the long term locking stability of our control system for about 17 hrs by monitoring the fluctuations in the error signal.
Locking stability (3$\sigma$) during this period was observed to be 0.42 MHz (3.0 MHz for 10 bit ADC, 16 hrs.) (see figure \ref{fig:long_term}, top panel). An Allan deviation \cite{allandev,howe2000total} for these measurements is shown in bottom panel of figure \ref{fig:long_term}.
These results show that for a 100 s averaging time, the stability goes up to \num{5e-11} and the overall stability is within \num{2e-10}. 
In these measurements, our feedback system was measuring the error signal at a rate of 5 Hz. 
In order to further suppress the noise, 5 such successive measurements were averaged.

\section{\label{sec:level1} Further possible improvements}

One simple way to further improve the stability of our He-Ne laser is by the optimization of proportional (P), integration (I), and derivative (D) values.
Currently, we have only optimized these values qualitatively by monitoring the error signal and further improvement might be possible.
Another point to be considered is to improve the stability of reference voltage.
Typically, we observe a 5 – 10 mV ripple ($V_{\rm pp}$, peak to peak voltage) in the reference voltage ($V_{\rm ref}$) of our microcontroller that was used to drive ($V_{\rm cc}$) a 16 bit ADC (see figure 1 (c)). 
The effect of this instability in the $V_{\rm ref}$ is reduced by  averaging to some extent.
However, a better solution would involve the use specialized ICs for providing a highly stable reference voltage (for example REF4132B50DBVR, 5-10 ppm stability).
Beside the above mentioned points, we also observe that the laser cavity was very sensitive to  back-scattered light. 
In our case, a small amount of back reflection from the fiber input face (P1-630A-FC-1) is enough to destabilize the laser and have a measurable influence on its locking stability. 
This can be minimized by small modifications in the experimental setup. For example in our case, we minimized it by coupling light into the fiber with a slight angular offset such that the back reflection is minimized. 
For complete elimination one has to use Faraday isolators or use FC/APC type connector. 
We quantified this disturbance by measuring the error signal with and without fiber coupling (see SI-7).

\section{Concluding Remarks}
We have successfully demonstrated frequency stabilization and characterization of He-Ne laser to ppb level stability, built using simple and low cost components.
For a measurement over 17 hours time span, our locked He-Ne laser has a stability of 0.42 MHz (3$\sigma$).
Such a device can be a very valuable frequency reference, especially in applications involving high-resolution laser spectroscopy.
Given the simple nature of the design, we also believe that this work has potential to provide valuable experience in feedback control systems as well as the learning fundamental concepts related to laser cavities, when included as part of undergraduate lab exercise.
Finally, this work also provides a preview into the potential of the wavemeter used in this work.
This simple and robust design can be extended to develop a stand alone high precision and accuracy, low-cost wavemeter, and is currently being pursued in our lab.

\section*{SUPPLEMENTARY INFORMATION}

\begin{itemize}
\item SI-1: Details of components used for building stabilized He-Ne laser
\item SI-2: Details of components used for the interferometer
\item SI-3: PID code for microcontroller   
\item SI-4: He-Ne laser cavity and Fizeau wedge characterization
\item SI-5: Mode hop free scanning under Neon gain profile
\item SI-6: Calibration of error signal (bits) in term of frequency unit (MHz) for 16 bit ADC
\item SI-7: Effect of back reflection on locking stability
\end{itemize}

\section*{AUTHOR CONTRIBUTIONS}

SKS designed the experiments, performed the measurements and data analysis with inputs from PRS. AK contributed to the experimental design and testing at initial stages of the work. PRS provided conceptual inputs and designed the project. SKS and PRS prepared the manuscript. All authors discussed the results and contributed to the manuscript.

\section*{Data Availability}
Files related to the experimental data presented in this manuscript and supplementary information can be accessed from the following link:
{https://doi.org/10.5281/zenodo.7152400}
\section*{COMPETING INTERESTS}
The authors declare no competing interests.
\begin{acknowledgments}
This work was supported by intramural funds at Tata Institute of Fundamental Research - Hyderabad, Department of Atomic Energy, India.
\end{acknowledgments}

\bibliography{ref}

\end{document}


\title{\textbf{A simple and low-cost setup for part per billion level frequency stabilization and characterization of red He-Ne laser -- Supplementary Information}}

\author{Saurabh Kumar Singh, Avinash Kumar, Pranav R. Shirhatti*}

\affil{Tata Institute of Fundamental Research Hyderabad,
36/P Gopanpally, Hyderabad 500046, Telangana, India\\ *\email{pranavrs@tifrh.res.in}}

\date{}

\maketitle

\section*{SI-1: Details of components used for building stabilized He-Ne laser}

 \begin{table*}[ht]
 \caption{Details of components used for building the stabilized He-Ne laser}\label{tab1}\vspace{.5cm}
\centering
\begin{tabular}{ |p{2cm}|p{2.3cm}|p{2.0cm}|p{1.7cm}|p{2.1cm}|p{2cm}|}
 
 \hline
S. no. & Component name & Part number & Quantity & Approximate cost
(USD) & Source\\
 \hline
 1   & He-Ne tube    &Melles Griot, 05-LHR-006&   1&   210& Meredith Instruments\\
 \hline
 2& He-Ne Laser Power Supply    & DG-22-00 &   1&   220& Meredith Instruments\\
 \hline
 3 & Polarizing beam splitter cube    &PBS102 &   1&   199& Thorlabs \\
 \hline
   4 & Si Photodiodes   &FDS-100 &   2&   22& Thorlabs\\
 \hline
 5 & Dichroic Film Polarizer Sheet    &LPVISE2X2 &   1&   9& Thorlabs\\
 \hline
 6 & Microcontroller    & Arduino Uno &   1&   6& \\
 \hline
 7 & OpAmp IC’s    &uA741C &   2&   1& \\
 \hline
  8 & MOSFET switch   &IRF520N &   4&   1& \\
 \hline
 9 & 16-Bit ADC    & ADS1115&   1&   4-5& \\
 \hline
  10 & 12 W power supply adapter   &ECA-12W-12 12V 1A  &   1&   2-3& \\
 \hline

 \multicolumn{4}{|l|}{Total estimated cost}&USD: 671& \\ \cline{1-4} 
 \hline
\end{tabular}
\end{table*}

\newpage

\section*{SI-2: Details of components used for  the interferometer}
 \begin{table*}[ht]
 \caption{Details of components used for building the interferometer used for characterizing He-Ne}\label{tab1}
\begin{center}
    
\begin{tabular}{ |p{2cm}|p{2.3cm}|p{2.0cm}|p{1.7cm}|p{2.1cm}|p{2cm}|}
 \hline
 Serial number& Component name &Part number&Quantity& Approximated cost
(USD) & Source\\
 \hline
 1   & Wedge window    &WW41050 – Ø1" UVFS Wedged Window&   1&   105& Thorlabs\\
 \hline
 2& Single mode fiber    & SM, FC/PC, 633-780nm, FT030-Y &   1&   74& Thorlabs\\
 \hline
 3 & Fiber coupler   &F230FC – B– 633 nm, f = 4.43 mm, NA = 0.56 FC/PC connecter &   1&   160& Thorlabs\\
 \hline
 4 & Webcam   &Quantum QHM495LM6 Webcam, 640×480 pixel &  1&   5& \\
 \hline
 \multicolumn{4}{|l|}{Total estimated cost}&USD: 345& \\ \cline{1-4} 
 \hline
\end{tabular}
\end{center}

\end{table*}

\newpage

\section*{SI-3: PID code for microcontroller}

For our feedback control, we have used PID library version 1.2.1 (Brett Beauregard,  \url{https://github.com/br3ttb/Arduino-PID-Library}). The variables used in this library are defined as given by Equation 1. 
In our case the error signal is defined as the difference of the signal of the normalized photo diode. For mode balance stabilization, the error signal must always be zero.
\begin{equation} \label{eq3}
    \text{output} = \ K_{p}e + \ K_{d}\frac{de}{dt} + \ K_{i}\int_{0}^{t} e(t) \,dt 
\end{equation}

Here, e = Setpoint - Input

 \begin{lstlisting}[language=C, caption=PID code for He-Ne stabalization]
//===========================================//
// PID CODE for HE-NE stabilization
//===========================================//
#include <PID_v1.h> //PID library
#include <Adafruit_ADS1X15.h>  //16 bit ADC library
Adafruit_ADS1115 ads;
 float Voltage = 0.0;
  float Voltage1 = 0.0;
const int numReadings =45;
#define PIN_INPUT lightLevel
//#define PIN_INPUT 4

#define RELAY_PIN 8
const int led = 8; // LED output
double lightLevel;
int readings[numReadings];
int readings1[numReadings];// the readings from the analog input
int readIndex = 0;              // the index of the current reading
float total = 0;                  // the running total
float average = 0;                // the average
float total1 = 0;                  // the running total
float average1 = 0;                // the average

float norm=0;
 float norm_pd1=0;
 float norm_pd2=0;
float err_sig=0;

double Setpoint, Input, Output;  //These are just variables for storing values
PID myPID(&Input, &Output, &Setpoint,4.6, .050551851154,.644488081098, DIRECT); // This sets up our PID Loop
//Input is our PV
//Output is our u(t)
//Setpoint is our SP
const int sampleRate = 1; // Variable that determines how fast our PID loop runs12
// Communication setup
const long serialPing =180;
unsigned long now = 0; //This variable is used to keep track of time
// placehodler for current timestamp
unsigned long lastMessage = -10; //

// seting the programmable gain amplifier (PGA)
void setup()
{
Serial.begin(9600);
//ads.setGain(GAIN_TWO);        +/- 2.048V  1 bit = 0.0625mV
//ads.setGain(GAIN_FOUR);      // +/- 1.024V  1 bit = 0.03125mV
//ads.setGain(GAIN_EIGHT);   //   +/- 0.512V  1 bit = 0.015625mV
ads.setGain(GAIN_SIXTEEN);    //+/- 0.256V  1 bit = 0.0078125mV  

ads.begin();

Setpoint =0;
myPID.SetMode(AUTOMATIC);  //Turn on the PID loop
  myPID.SetSampleTime(sampleRate); //Sets the sample rate
 lastMessage = millis();
}
 
void loop()
{
int16_t adc0, adc1, adc2, adc3;
  total = total - readings[readIndex];
  total1 = total1 - readings1[readIndex];
  // read from the sensor:
 readings[readIndex] = ads.readADC_SingleEnded(2);
  readings1[readIndex] = ads.readADC_SingleEnded(3);
  // add the reading to the total:
  total = total + readings[readIndex];
    total1 = total1 + readings1[readIndex];

  // advance to the next position in the array:
  readIndex = readIndex + 1;

  // if we're at the end of the array...
  if (readIndex >= numReadings) {
    // ...wrap around to the beginning:
    readIndex = 0;
  }

  // calculate the average:
  average = total / numReadings;
  average1 = total1 / numReadings;
//adc0 = ads.readADC_SingleEnded(0);
Voltage =(average * 0.1875)/1000;
Voltage1 =(average1 * 0.1875)/1000;

norm = (Voltage+Voltage1);
  norm_pd1=(Voltage/norm)*50000;
  norm_pd2=(Voltage1/norm)*50000;
  err_sig=(norm_pd2-norm_pd1);

int Setpoint =0; //Read our setpoint
lightLevel = err_sig;
  
  //lightLevel = voltageA2; //Get the light level
  Input = lightLevel; //Map it to the right scale
  
  myPID.Compute();  //Run the PID loop
  analogWrite(8, Output); 
now = millis(); //Keep track of time
  if(now - lastMessage > serialPing) { 
//Serial.print("AIN0: ");

Serial.print( err_sig,5);

Serial.print(" ");

Serial.print(norm_pd2,5);

Serial.print(" ");
Serial.print(norm_pd1,5);

Serial.println(" ");
delay(.05);
lastMessage = now; 
}}
\end{lstlisting}

\section*{SI-4: He-Ne laser cavity and Fizeau wedge characterization}
In this section, some additional details of the calibration of the interferometer in order to establish a relation among fringe position (in pixels), frequency and FSR of the wedge.

\subsection*{Wedge spacing calibration with Dye laser}

\begin{figure}[H]
\centering
\includegraphics[width=16cm]{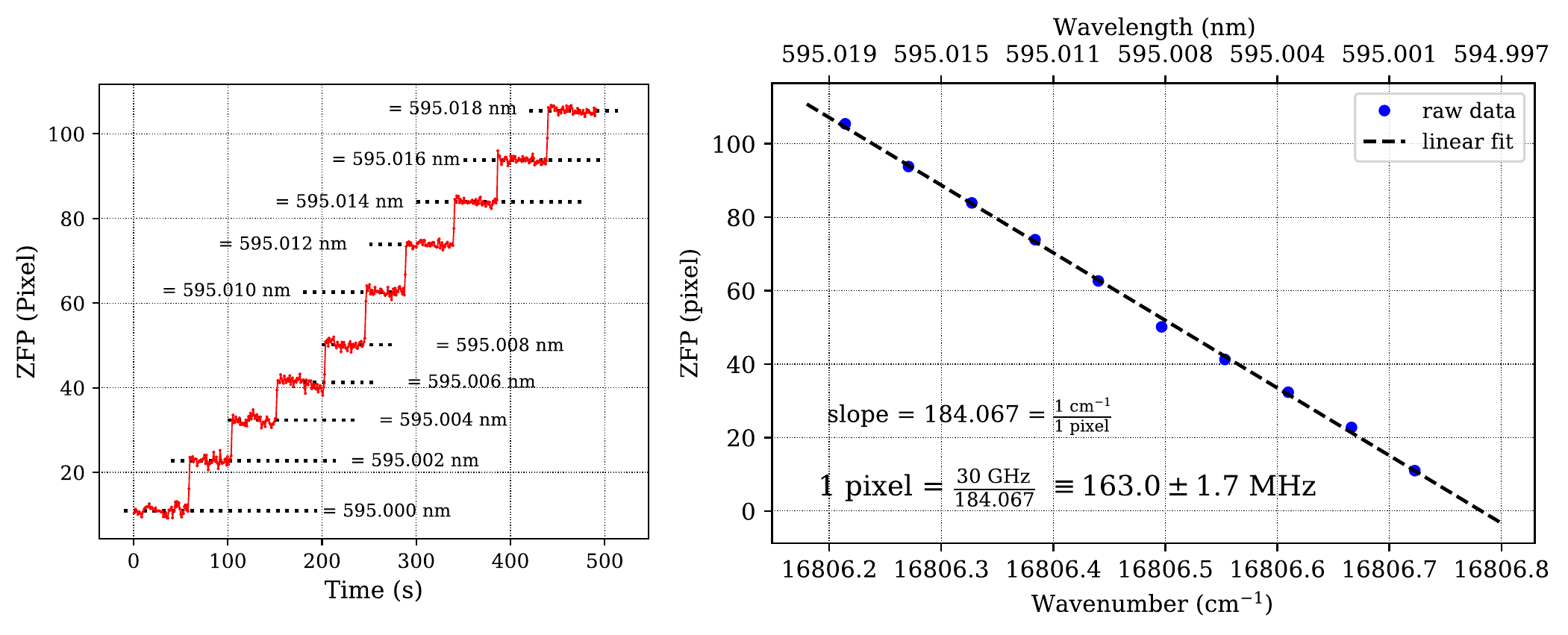}
\caption{Left panel shows how Zeroth fringe position (ZFP) changes over wavelength with step size of 0.002 nm. In right panel, mean values of ZFP are plotted verses wavelength to get relation among ZFP shift (in pixel) and frequency.  }\label{fig:dye_laser_method}
\end{figure}

FSR of the wedge was estimated to be 20.2 GHz based on a measured physical thickness of 4.8 $\pm$ 0.1 mm (equation \ref{eq:fsr}). 
The refractive index of UV fused silica wedge ($n_{w}$) is 1.45717 at 632.8 nm and $e_{s}$ is the wedge thickness.
This FSR value is equal to the separation between two consecutive maxima or minima of interference pattern which is 129.30 pixels [see manuscript figure 3(c)].
\begin{equation} \label{eq:fsr}
    \ FSR_{w} = \frac{c}{2\ n_{w}\ e_{s}} =129.30 \hspace{.2 cm}\text{pixel} 
\end{equation}
For a more accurate measurement of FSR, we measured wedge spacing using a tunable dye laser.
Our dye laser has minimum step size of 0.0001 nm and it is independently calibrated with iodine absorption spectrum. 
The changes in zeroth fringe position (ZFP) were measured by scanning the wavelength (with 0.002 nm step
size) and is shown in figure \ref{fig:dye_laser_method} (left).

Mean value of each measured ZFP distribution were plotted verses corresponding wavelength, as shown in figure \ref{fig:dye_laser_method} (right). 
From the slope of the linear fit we established a relation between change in ZPF (in pixel) and frequency. The FSR of fizeau wedge was calculated to be 21.1 $\pm$ 0.2 GHz (equation \ref{eq:fsr}).


\subsection*{Determining FSR of the wedge using the longitudinal modes of He-Ne}

The spacing between adjacent longitudinal modes of our He-Ne is 1078 MHz (as per data sheet).

\begin{figure}[H]
  \centering
  \includegraphics[width=16cm]{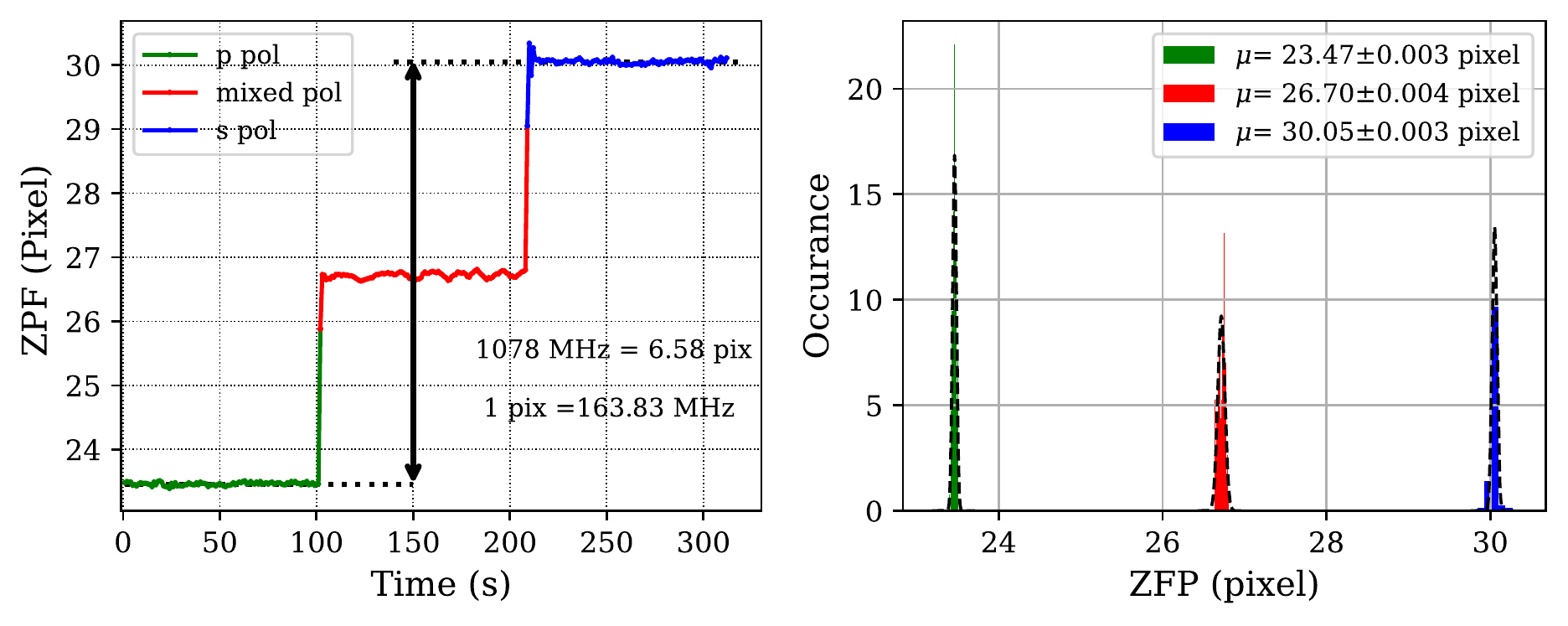}
  \caption{(Left) ZFP (in pixels) measured for single as well as mixed polarization using our interferometer.(right) Histogram plot of the ZFP measured for S, mixed polarization, and P polarization, respectively.}
  \label{fig:zfpvspol}
\end{figure}

Using this fact we estimated the wedge spacing by monitoring ZFP for different polarization light in the interferometer. 
We use a Polarizing beam splitter cube (1000:1, mode discrimination)  
which allows select a single polarization to enter the interferometer at given time.
The total change in ZFP for changing from S to P polarization was observed to be 6.58 pixel, as shown in figure \ref{fig:zfpvspol} (left). Histogram of the measured ZPF distribution over time, for S, mixed and P polarization, are shown in \ref{fig:zfpvspol} (right). 

The standard error of the mean (SEM) in the measured ZFP distributions for the S, mixed and P polarizations was measured to be 0.003, 0.004 and 0.003 pixels, respectively. The measured SEM values showed that the interferometer had the ability to determine the change in ZPF better than 0.005 pixels. Having said that the thermal drift dominates at the longer time scales of 15 min [see manuscript figure 4(a)].
As a conservative estimate, we  consider the interferometer drift of 0.02 pixel for 15 min as the overall uncertainty.
Given that the frequency spacing between the adjacent modes is 1078 MHz, a relation among frequency shift and the fringe position deviation is readily obtained
as 1 pixel = 163.83 MHz. 
Hence, FSR of the wedge was determined to be 21.183 $\pm$ 0.005 GHz.


\begin{table}[h]


 \caption{ Calculated wedge spacing (mm) and its FSR (GHz) from estimated using different methods}\label{tab:fsr}\vspace{.5cm}

\centering 
\begin{tabular}{l  c  c   rrr} 
\hline\hline 
Method & Calibration reference & wedge spacing (mm) & FSR of wedge (GHz)
\\ [0.5ex]
\hline \\

Physical thickness measurement & 0.1 mm least count vernier caliper &  4.8&  21.4 $\pm$ 0.4  \\[1ex]
Method 1 &Tunable dye laser&  4.87&  21.1 $\pm$ 0.2  \\[1ex]


Method 2 &Polarizing beam splitter cube  &4.857 &   21.183 $\pm$ 0.005 \\[1ex]

\hline 
\end{tabular}
\end{table}

The methods described above (using a tunable dye laser source or a polarizing beam splitter cube) gave almost identical wedge spacing values and the solid wedge has an FSR of about 21.1 - 21.3 GHz (see table \ref{tab:fsr}).
Having said that method 2 was more accurate and was also verified with dye laser, the result was also consistent with He-Ne tube manufacturing datasheet. For calibration purposes we used the method 2.

\section*{SI-5: Mode hop free scanning under Neon gain profile}

In general, He-Ne laser is used as a single wavelength laser. 
However, the feedback system implemented by us allows us to tune its wavelength in a controlled manner.
To verify tuning and mode hop scanning, we were measured the fringe pattern from single polarization (using polarizing beam splitter cube) in the interferometer. 
The ZFP was measured for a passively controlled cavity, and mode hop free scan was observed, as indicated by the fringe position shift (figure \ref{fig:mhfscan}). 
This can also be done actively using the feedback system to obtain an electronically controlled scan.

\begin{figure}[H]
\centering
\includegraphics[width=8cm]{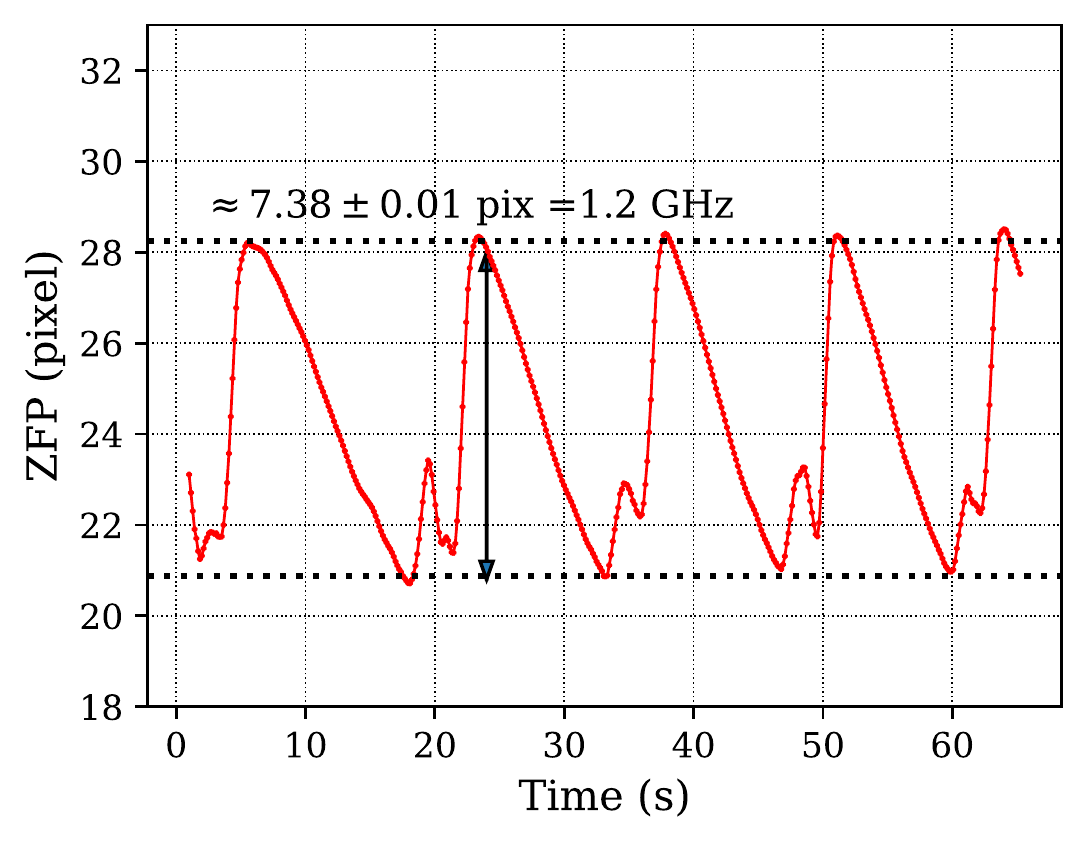}
\caption{ZFP verses time for a single polarization of He-Ne. In these measurements active feedback was turned off and the cavity was slowly allowed to drift (thermally).}
\label{fig:mhfscan}
\end{figure}

A total tuning was 1.2 $\pm$ 0.005 GHz observed, which is consistent with the expected gain bandwidth of the He–Ne laser.
\section*{SI-6: Calibration of error signal (bits) in terms of
frequency unit (MHz) for 16 bit ADC}
\begin{figure}[H]
\centering
\includegraphics[width=16cm]{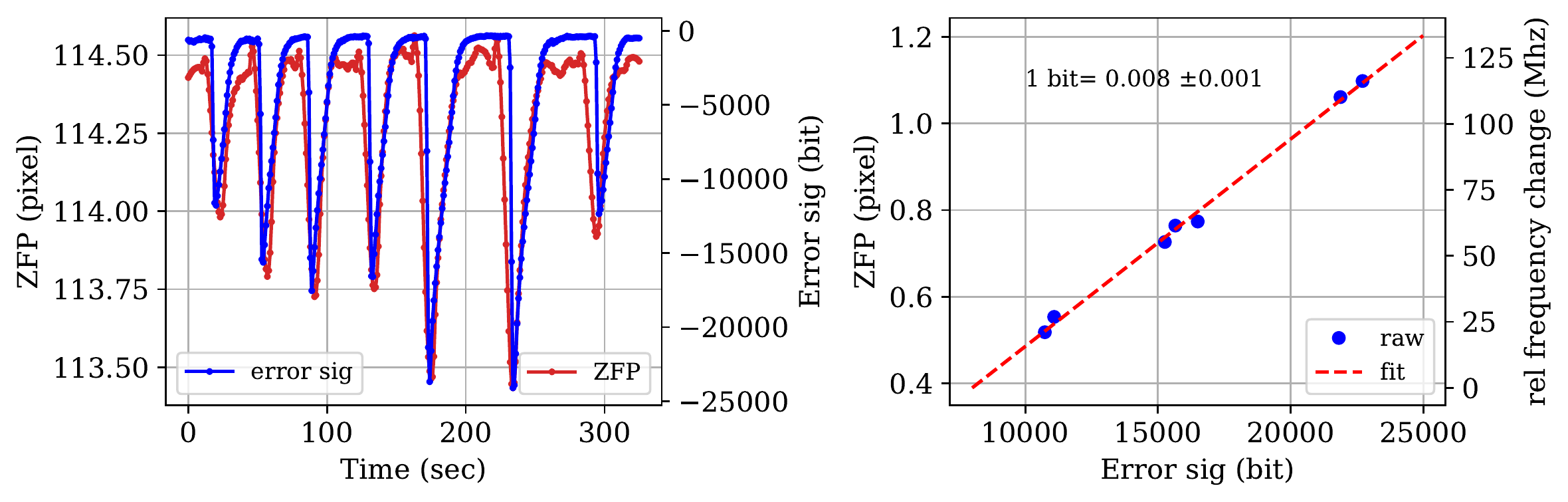}
\caption{Calibration of error signal (in digital units) to frequency units (in MHz). (Left) Changes in ZFP values were plotted with the error signal being changed independently by external disturbance (created by blowing cold air for a few seconds). (Right) Changes in both the error signal and the ZFP value were fitted with the linear equation. Based on the best fit parameters (1 pixel = 163.18 MHz) we estimate that change in error signal by 1 bit corresponds to a change in frequency of 0.008 $\pm$ 0.001 MHz .}
\label{fig:mhfscan}
\end{figure}

\section*{SI-7: Effect of back reflection on locking stability}
We quantified this disturbance by measuring the error signal with and without fiber coupling which is shown in figure 5. We found that
the stability of the He-Ne laser is almost 2-3 times poor
when it ics coupled using the fiber, compared to free space
coupling.
\begin{figure}[H]
\centering
\includegraphics[width=8cm]{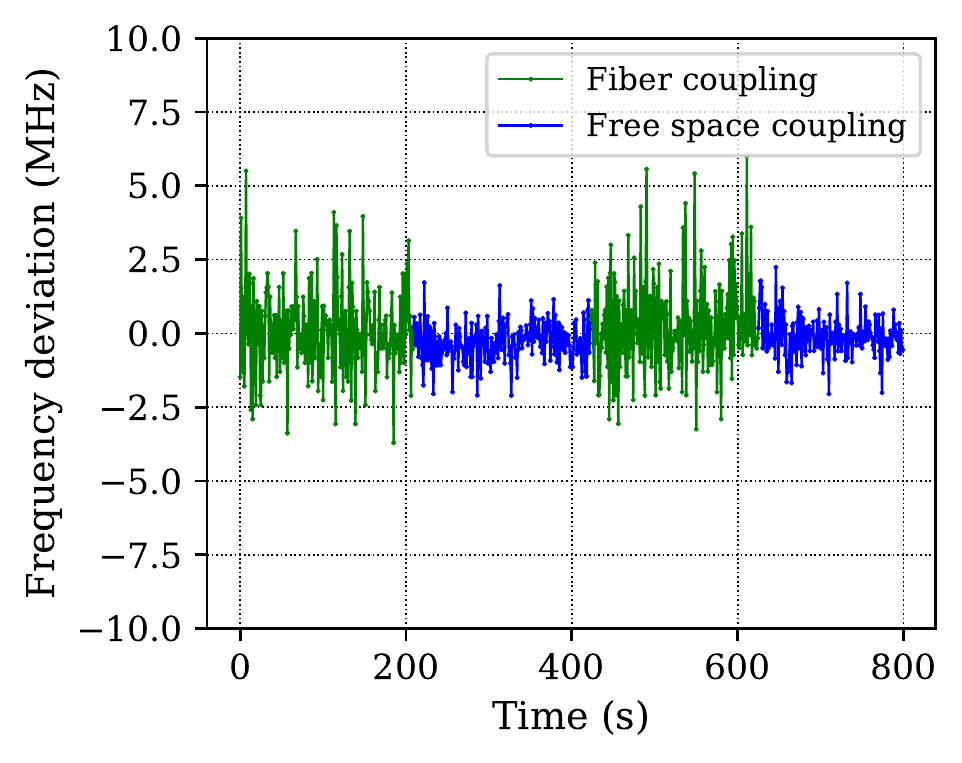}
\caption{Frequency drift observed with and without fiber coupling. Frequency instability was observed to increase by a
factor of 2-3 (caused back reflection). These measurements were made with the 10-bit ADC.}
\label{fig:mhfscan}
\end{figure}